\documentclass[aps,prd,longbibliography,twocolumn,showpacs,amsmath,amssymb,superscriptaddress,nofootinbib,floatfix,10pt,preprintnumbers,groupedaddress]{revtex4-1}

\usepackage{graphicx}
\usepackage{dcolumn}
\usepackage{bm}
\usepackage{gensymb}
\usepackage{verbatim}

\usepackage[usenames,dvipsnames]{color}
\include{jdefs}
\usepackage{multirow}
\usepackage{url}
\usepackage{xspace}
\usepackage{braket}
\usepackage{tabularx}
\definecolor{darkblue}{rgb}{0,0,0.5}
\definecolor{darkgreen}{rgb}{0.1,0,0.3}
\definecolor{darkred}{rgb}{0.6,0,0}
\usepackage[bookmarks=true, bookmarksnumbered=true, colorlinks=true,
  urlcolor=darkblue, citecolor=darkgreen, linkcolor=darkred,
  breaklinks=true]{hyperref} 
\usepackage[normalem]{ulem}

\begin{document}

\preprint{KCL-PH-TH/2017-63}

\title{Prospects for detecting eV-scale sterile neutrinos from a galactic supernova}

\author{Tarso Franarin}
\email{tarso.franarin@kcl.ac.uk}
\author{Jonathan H. Davis}%
\email{jonathan.davis@kcl.ac.uk}
\author{Malcolm Fairbairn}%
\email{malcolm.fairbairn@kcl.ac.uk}
\affiliation{Theoretical Particle Physics and Cosmology, Department of Physics,
King's College London, London WC2R 2LS, United Kingdom}

\date{\today}

\begin{abstract}
Future neutrino detectors will obtain high-statistics data from a nearby core-collapse supernova. We study the mixing with eV-mass sterile neutrinos in a supernova environment and its effects on the active neutrino fluxes as detected by Hyper-Kamiokande and IceCube. Using a Markov Chain Monte Carlo analysis, we make projections for how accurately these experiments will measure the active-sterile mixing angle $\theta_s$ given that there are substantial uncertainties on the expected luminosity and spectrum of active neutrinos from a galactic supernova burst. We find that Hyper-Kamiokande can reconstruct the sterile neutrino mixing and mass in many different situations, provided the neutrino luminosity of the supernova is known precisely. Crucially, we identify a degeneracy between the mixing angle and the overall neutrino luminosity of the supernova. This means that it will only be possible to determine the luminosity if the presence of sterile neutrinos with $\theta_s \gtrsim 0.1^{\circ}$ can be ruled out independently. We discuss ways in which this degeneracy may be broken in the future.

\end{abstract}

\maketitle


\section{Introduction}
When a star with a mass greater than approximately eight solar masses reaches the end of its life, it collapses under its own gravity, leaving behind a neutron star or a black hole. This collapse is called a supernova (SN), and leads to the production of as many as $10^{57}$ neutrinos in  around ten seconds, some of which come from electron capture onto protons but the majority of which are produced thermally from the hot core of the collapsing star \cite{Mirizzi:2015eza}. Three decades ago SN1987A provided the only detection of neutrinos from a core-collapse supernova to date, despite the fact that the explosion took place outside the Galaxy in the Large Magellanic Cloud at a distance of around 50 kpc and that neutrino detectors were then in their relative infancy. We expect at least a few galactic supernovae per century \cite{Ando:2005ka}, and so when the next one occurs, large neutrino detectors like Super-Kamiokande (or Hyper-Kamiokande) \cite{Abe:2011ts,Hyper-Kamiokande:2016dsw} and IceCube \cite{Aartsen:2014gkd} will be able to record high-statistics data of the supernova neutrino spectrum \cite{Horiuchi:2017sku,Nikrant:2017nya}.

All predictions for the spectrum of supernova neutrinos come from simulations \cite{Horiuchi:2017qja}. These depend sensitively on the complicated physics of a core-collapse event, and also on the properties of the progenitor star, introducing uncertainty into our expectation of quantities such as the average neutrino energy and the total neutrino luminosity. These simulations also depend on neutrino physics, for example the ordering of neutrino masses and the potential existence of additional neutrino species, called sterile neutrinos, which mix only a tiny amount with the known active flavours and do not interact with matter. It is the effect of these sterile neutrinos which we focus on in this work.

Short baseline neutrino oscillation experiments such as LSND \cite{Aguilar:2001ty}, MiniBooNE \cite{AguilarArevalo:2008rc,Aguilar-Arevalo:2013pmq}, reactor experiments \cite{Mention:2011rk}, and gallium source experiments \cite{Abdurashitov:2009tn,Kaether:2010ag,Giunti:2012tn} may hint at the existence of a third mass splitting on the eV-scale. Since LEP Z$^0$ decay measurements are consistent with only three active neutrinos \cite{ALEPH:2005ab}, an additional light fourth neutrino flavor must be sterile. Other similar experiments have observed null results. Global fits employing three active neutrinos and one sterile neutrino, called 3+1 models, are still able to accommodate all available data \cite{Gariazzo:2017fdh}. New short baseline experiments will give a definitive answer on the existence of sterile neutrinos connected with these anomalies \cite{Borexino:2013xxa,Harada:2013yaa,Bhadra:2014oma,Antonello:2015lea,Ciuffoli:2015uta,Gollapinni:2015lca,Axani:2015zxa,Abs:2015tbh,Ashenfelter:2015uxt,Barinov:2016znv,Michiels:2016qui,Alekseev:2016llm,Serebrov:2017nxa,Manzanillas:2017rta}.

The existence of a sterile neutrino state could have significant effects on a SN environment such as enhancing the neutron abundance and allowing the production of heavy elements \cite{Tamborra:2011is,Wu:2013gxa} or suppressing the neutronization burst for the inverted mass ordering \cite{Esmaili:2014gya}. In this paper we investigate how an additional light sterile neutrino would change the observed neutrino flux and spectrum from a future nearby supernova in Hyper-Kamiokande and IceCube, given that we do not know precisely the expected spectrum for the active neutrinos even in the case where sterile neutrinos do not exist.

We make projections for how well experiments, such as Hyper-Kamiokande or IceCube, will be able to constrain or measure the mixing angle between sterile and active neutrino states, after observing the neutrino burst from a galactic supernova. We focus on the effect of our uncertain knowledge of the spectrum of the active flavours from a supernova burst on our ability to constrain such sterile states.
In the next section we will describe the production of neutrinos in supernovae as well as their mixing and behaviour as they leave the star.  We explain the assumptions we have made and present some consistency checks  with regards to our approach.  We then consider the detection of these neutrinos once they arrive at Earth with the proposed detector Hyper-Kamiokande and the Ice Cube detector at the South Pole.  Finally  we discuss what can be learned about sterile neutrinos with such observations.

\section{Neutrino production and flavour conversion in supernovae}
During a supernova core-collapse, the large amount of gravitational energy released is enough to heat the stellar matter to temperatures exceeding 10~MeV.
Neutrinos are produced during collapse both by beta and thermal processes \cite{Janka:2017vlw}. The most important beta-processes are electron capture by nuclei or free protons, which occur early in the supernova lifetime during the neutronisation burst:
\begin{equation}
\begin{split}
&e^-+(A,Z)\rightarrow (A,Z-1)+\nu_e,\\
&e^-+p\rightarrow n+\nu_e.
\end{split}
\label{eq:beta}
\end{equation}
This is followed by thermal production of all neutrino flavours.
Significant thermal emission processes are pair annihilation of $e^+e^-$ pairs, nucleon-nucleon bremsstrahlung and plasmon decay:
\begin{equation}
\begin{split}
&e^-+e^+\rightarrow\nu+\bar{\nu},\\
&N+N\rightarrow N+N+\nu+\bar{\nu},\\
&\text{(plasma excitation)} \rightarrow \nu+\bar{\nu}.
\end{split}
\label{eq:thermal}
\end{equation}

Neutrinos propagating in matter are subject to a potential due to the coherent forward elastic scattering with the particles in the medium \cite{Wolfenstein:1977ue}:
\begin{equation}
\begin{split}
&\nu+N\rightarrow\nu+N,\\
&\nu+e^-\rightarrow\nu+e^-\\
&\nu+\nu\rightarrow\nu+\nu.
\end{split}
\label{eq:scatterings}
\end{equation}
These interactions give rise to a potential for each neutrino flavour \cite{Notzold:1987ik}:
\begin{equation}
\begin{split}
&V_{\nu_e}(r)=\sqrt{2}G_F(N_e-0.5N_n+2N_{\nu_e}),\\
&V_{\nu_x}(r)=\sqrt{2}G_F(-0.5N_n+N_{\nu_e}).
\end{split}
\label{eq:potentials}
\end{equation}
Here $N_e$, $N_n$ and $N_{\nu_e}$ are the number densities of electrons, neutrons, and electron neutrinos minus
the number densities of their antiparticles, respectively. The subscript $x$ stands for both $\mu$ and $\tau$ flavours. The effective potential for sterile neutrinos is zero and for antineutrinos $V_{\bar{\nu}_\alpha} = - V_{\nu_\alpha}$. Since $\mu$ and $\tau$ neutrinos and antineutrinos are produced exclusively in equal quantities by thermal pair production in a supernova there are no net contributions to equation (\ref{eq:potentials}) that depend on $N_{\nu_\mu}$ and $N_{\nu_\tau}$.

The flavour neutrino states, $\nu_f\equiv(\nu_e,\nu_\mu,\nu_\tau,\nu_s)^T$ evolve according to
\begin{equation}
i\frac{d\nu_f}{dr}=H\nu_f=(H_0+V)\nu_f=(U\frac{M^2}{2E_\nu}U^\dagger+V)\nu_f.
\label{eq:evolution}
\end{equation}
Here $U=R_{34}R_{24}R_{14}R_{23}R_{13}R_{12}$ is the 3+1 mixing matrix where each $R_{ij}$ is a rotation matrix through angle $\theta_{ij}$ in the $ij$ plane, $M^2=\text{diag}(m_1^2,m_2^2,m_3^2,m_4^2)$ is the mass matrix and $V=\text{diag}(V_{\nu_e},V_{\nu_\mu},V_{\nu_\tau},V_{\nu_s})$ is the potential matrix. 

Interactions in matter modify the mixing of neutrinos, which can be large even if the mixing angle in vacuum is small. At any instant the effective Hamiltonian $H$ in equation (\ref{eq:evolution}) can be diagonalised by a unitary transformation and the neutrino propagation is then described with respect to the instantaneous mass eigenstates of the Hamiltonian in matter $\nu_m\equiv(\nu_{1m},\nu_{2m},\nu_{3m},\nu_{4m})^T$:
\begin{equation}
\begin{split}
\nu_f&=U^m\nu_m,\\
U^{m\dagger}HU^m=H^\text{diag}&=\text{diag}(E_1,E_2,E_3,E_4).
\end{split}
\end{equation}
where $E_i$ are the eigenvalues of the Hamiltonian. The matrix $U^m$ has the same form as $U$ but the vacuum mixing angles are replaced by the mixing angles in matter. Thus in a varying density medium the mixing angles change at every instant. 

The evolution equation can be written in the basis of the instantaneous eigenstates as
\begin{equation}
i\frac{d\nu_m}{dr}=\left[H^\text{diag}-iU^{m\dagger}\frac{dU^m}{dr}\right]\nu_m.
\label{eq:evolution_m}
\end{equation}
If density changes slowly enough then the transitions between instantaneous eigenstates $\nu_{im}$ are suppressed and the change of flavour follows the density variations i.e. we have adiabatic conversion \cite{Akhmedov:1999uz}. This is the MSW effect \cite{Mikheev:1986wj,Mikheev:1987jp}, which was used to successfully explain the solar neutrino problem \cite{Bethe:1986ej}. 

Whenever two diagonal terms of the Hamiltonian written in the flavour basis (equation (\ref{eq:evolution})) are equal, there is a resonance. At these points the difference between the energies of the mass eigenstates is minimised and the propagation reaches its lowest adiabaticity. If the propagation is adiabatic at the resonance, it will then also be adiabatic elsewhere. To quantify the adiabaticity at resonance we assume a $2\times 2$ factorization of dynamics and use the Landau-Zener approximation \cite{landau,zener}.  In this approximation, the probability for transitions between two eigenstates $i$ and $j$ (``jumping probability'') is:
\begin{equation}
P_{\text{jumping}}=\text{exp}\left[-\frac{\pi^2}{2}\frac{\Delta r}{l_m^{res}}\right],
\label{eq:pjumping}
\end{equation}
with resonance width
\begin{equation}
\Delta r = 2\,\text{tan}2\theta_0\left|\frac{1}{V}\frac{\partial V}{\partial r}\right|^{-1}
\end{equation}
and oscillation length at resonance 
\begin{equation}
l_m^{res} = \frac{2\pi}{|E_i-E_j|_{res}}.
\end{equation}
In the adiabatic limit this probability $P_{\text{jumping}}$ is zero. Table \ref{tab:resonances} lists the resonances that can occur in a SN environment due to active-active and active-sterile mixings.

\begin{table}[h]
\centering
\begin{tabular}{c|c}
Resonance & Condition  \\ \hline
$R_{e\mu}$ & $H_0^{\mu\mu}-H_0^{ee}=V_{\nu_e}(r)-V_{\nu_x}(r)$ \\
$R_{e\tau}$ & $H_0^{\tau\tau}-H_0^{ee}=V_{\nu_e}(r)-V_{\nu_x}(r)$ \\
$R_{es}$ & $H_0^{ss}-H_0^{ee}=V_{\nu_e}(r)$\\
$R_{\mu s}$ & $H_0^{ss}-H_0^{\mu\mu}=V_{\nu_x}(r)$\\
$R_{\tau s}$ & $H_0^{ss}-H_0^{\tau\tau}=V_{\nu_x}(r)$
\end{tabular}
\caption{The different resonances that can occur in a supernova due to active-active and active-sterile mixings.}
\label{tab:resonances}
\end{table}

We have used the matter densities from the simulation of an 8.8 $M_\odot$ supernova by the Garching group \cite{Huedepohl:2009wh}, under the assumption that the supernova occurs $10$~kpc from Earth. These data provide the doubly differential neutrino distribution in energy and time without any  adiabatic conversion. The differential flux of neutrinos of a given flavour is given by:
\begin{equation}
F_\nu(E_\nu,t)=\frac{dN_\nu}{dt}f(E_\nu),
\end{equation}
where
\begin{equation}
f(E_\nu)=\frac{1}{\left<E_\nu\right>}\frac{(1+\alpha)^{1+\alpha}}{\Gamma(1+\alpha)}\left(\frac{E_\nu}{\left<E_\nu\right>}\right)^\alpha\text{exp}\left[-(1+\alpha)\frac{E_\nu}{\left<E_\nu\right>}\right]
\end{equation}
is the normalised energy spectrum ($\int f(E_\nu)dE_\nu=1$) with $\alpha=2\left<E_\nu\right>^2-\left<E_\nu^2\right>/(\left<E_\nu^2\right>-\left<E_\nu\right>^2)$, $\left<E_\nu \right>$ is the average neutrino energy and ${dN_\nu}/{dt}$ it the total flux. These simulated data files include the radial profiles of properties such as density, particle abundances and average neutrino energies in 0.1 second intervals.  At each resonance we use the simulated profiles to calculate the jumping probability (\ref{eq:pjumping}). 

\begin{figure}[t]
\includegraphics[width=.95\linewidth]{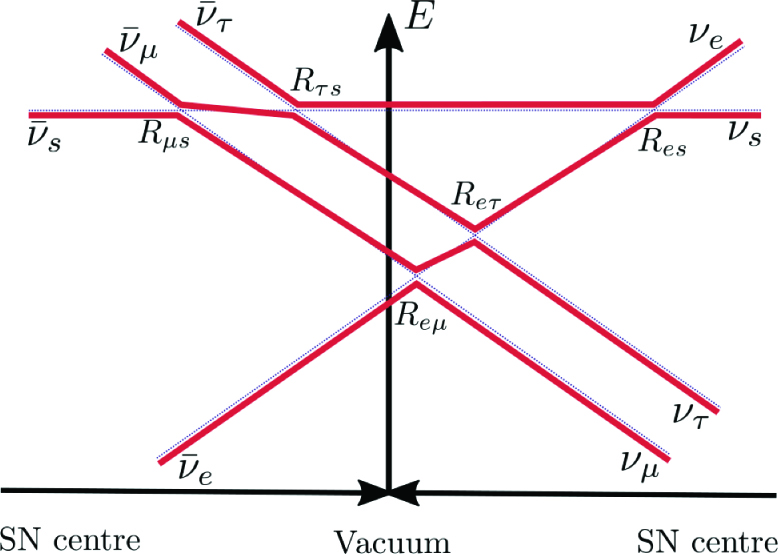}
\caption{The energies of the mass eigenstates in matter for normal mass ordering. The right-hand (left-hand) side of this diagram represents neutrinos (antineutrinos). In the production region flavour eigenstates coincide with the mass eigenstates. Propagation is adiabatic so the flavour content of each eigenstate follows the density variations and there are no transitions between eigenstates at resonances.}
\label{fig:diagram}
\end{figure}

At small radii, where most of the neutrinos are produced, the matter potential, which is diagonal in the flavour basis, dominates and so the flavour eigenstates coincide with the mass eigenstates. During the adiabatic propagation the flavour composition of each mass eigenstate follows the density variations. Figure \ref{fig:diagram} shows the eigenvalues $E_i$ of the mass eigenstates in matter for normal mass ordering as neutrinos propagate from the production region at small radii to the vacuum of empty space. In this figure we assume that all resonances in Table \ref{tab:resonances} occur and are adiabatic. The right-hand (left-hand) side of figure \ref{fig:diagram} represents neutrinos (antineutrinos). For example, the electron neutrinos produced in the inner regions of a supernova are essentially composed of the mass eigenstate $\nu_{4m}$. Since the propagation is adiabatic, there is no transition between mass eigenstates at the resonance $R_{es}$ and the flavour content of $\nu_{4m}$ follows the changes in density until it is mostly sterile in vacuum.

Neutrinos are produced in a coherent combination of mass eigenstates. Over a distance $L$, the separation between two wavepackets corresponding to two mass eigenstates with a given energy $E_\nu$ and mass difference $\Delta m^2$ due to different group velocities is \cite{Dighe:1999bi}:
\begin{equation}
\Delta L = \Delta v L = \frac{\Delta m^2}{2E_\nu^2} L.
\label{eq:coherence_loss}
\end{equation}
The size of an individual wavepacket can be estimated as $\sigma\lesssim 1/T \sim 10^{-11}$ cm. For $\Delta m^2 \sim 1$ eV$^2$ and $E_\nu\sim 10$ MeV the neutrinos lose coherence after a distance $L\sim 10$ m, still inside the star. For $\Delta m^2 \sim 10^{-5}$ eV$^2$ and $E_\nu\sim 10$ MeV the loss of coherence happens approximately at $L\sim 10^6$ m, in the outskirts of the stellar matter. This means that SN neutrinos propagate without coherence to Earth so oscillations in vacuum are irrelevant. The fluxes of neutrinos in the flavour basis at Earth are then given by
\begin{equation}
F_{\nu_\alpha}=\sum_{i=1}^4|U_{\alpha i}|^2F_{\nu_i},
\label{eq:flux_earth}
\end{equation}
with $F_i$ the mass eigenstates fluxes and $U_{\alpha i}$ the extended neutrino mixing matrix.

As an example we now show the calculation for antineutrinos in the normal ordering case. At small radii, where the majority of neutrinos are produced, the fluxes in the mass basis (left-hand side) are related to fluxes in the flavour basis (right-hand side) as
\begin{equation}
\begin{pmatrix}
F_{\bar{\nu}_1}^0 \\
F_{\bar{\nu}_2}^0 \\
F_{\bar{\nu}_3}^0 \\
F_{\bar{\nu}_4}^0 \\
\end{pmatrix}
=
\begin{pmatrix}
F_{\bar{\nu}_e}^0 \\
F_{\bar{\nu}_s}^0 \\
F_{\bar{\nu}_\mu}^0 \\
F_{\bar{\nu}_\tau}^0 \\
\end{pmatrix}.
\end{equation}
We naturally set $F_{\bar{\nu}_s}^0=0$ as there is no initial flux of sterile neutrinos. On their way out of the stellar matter all antineutrino mass eigenstates but $\bar{\nu}_{1m}$ pass through resonances. At the resonance $R_{\mu s}$ there is a probability $p_{R_{\mu s}}$ that the transition $\bar{\nu}_{2m}\leftrightarrow\bar{\nu}_{3m}$ occurs:

\begin{equation}
\begin{pmatrix}
F_{\nu_1}^\prime \\
F_{\nu_2}^\prime  \\
F_{\nu_3}^\prime  \\
F_{\nu_4}^\prime  \\
\end{pmatrix}
=
\begin{pmatrix}
1 & 0 & 0 & 0 \\
0 & 1-p_{R_{\mu s}} & p_{R_{\mu s}} & 0\\
0 & p_{R_{\mu s}} & 1-p_{R_{\mu s}} & 0 \\
0 & 0 & 0 & 1 \\
\end{pmatrix}
\begin{pmatrix}
F_{\bar{\nu}_e}^0 \\
F_{\bar{\nu}_s}^0 \\
F_{\bar{\nu}_\mu}^0 \\
F_{\bar{\nu}_\tau}^0 \\
\end{pmatrix}.
\end{equation}
After that a similar situation takes place at resonance $R_{\tau s}$, where the transition $\bar{\nu}_{3m}\leftrightarrow\bar{\nu}_{4m}$ may happen with probability $p_{R_{\tau s}}$:
\begin{equation}
\begin{pmatrix}
F_{\nu_1} \\
F_{\nu_2} \\
F_{\nu_3} \\
F_{\nu_4} \\
\end{pmatrix}
=
\begin{pmatrix}
1 & 0 & 0 & 0 \\
0 & 1 & 0 & 0\\
0 & 0 & 1-p_{R_{\tau s}} & p_{R_{\tau s}} \\
0 & 0 & p_{R_{\tau s}} & 1-p_{R_{\tau s}} \\
\end{pmatrix}
\begin{pmatrix}
F_{\nu_1}^\prime \\
F_{\nu_2}^\prime  \\
F_{\nu_3}^\prime  \\
F_{\nu_4}^\prime  \\
\end{pmatrix}.
\end{equation}

In the adiabatic limit all jumping probabilities are zero and from equation (\ref{eq:flux_earth}) the $\bar{\nu}_e$ flux on Earth, which is most relevant to Hyper-Kamiokande and IceCube, becomes 
\begin{equation}
F_{\bar{\nu}_e}=|U_{e1}|^2F_{\bar{\nu}_e}^0+|U_{e3}|^2F_{\bar{\nu}_\mu}^0+|U_{e4}|^2F_{\bar{\nu}_\tau}^0
\end{equation}
instead of 
\begin{equation}
F_{\bar{\nu}_e}=|U_{e1}|^2F_{\bar{\nu}_e}^0+|U_{e2}|^2F_{\bar{\nu}_\mu}^0+|U_{e3}|^2F_{\bar{\nu}_\tau}^0
\end{equation}
in the case without mixing with sterile neutrinos. The mixings $\theta_{14}=\theta_{24}=\theta_{34}=10\degree$ would imply a 30\% decrease on the flux.

The calculation of the neutrino flux, in the presence of sterile flavours, can be complicated by feedback effects. The electron abundance is set by the following reactions and its reverse processes:
\begin{align}
\nu_e+n\rightarrow p+e^-\\
\bar{\nu}_e+p\rightarrow n+e^+
\end{align}
At resonance $R_{es}$ electron neutrinos are converted to sterile neutrinos. The conversion probability depends on $N_e=n_{e^-}-n_{e^+}$ and its derivative at the resonance position. The depletion of electron neutrinos and the unchanged number of electron antineutrinos lower $N_e$ for $r>r_{R_{es}}$ and the potential gradient becomes steeper. Therefore the $R_{es}$, $R_{e\mu}$ and $R_{e\tau}$ conversions, which explicitly depend on $N_e$, become less adiabatic \cite{Nunokawa:1997ct}. This effect also increases the neutron abundance, which can be important for the propagation of antineutrinos. However, the resonances $R_{\mu s}$ and $R_{\tau s}$ take place before $R_{es}$ so we can safely neglect feedback effects for antineutrinos. Since the main detection channel on Hyper-Kamiokande and IceCube is electron antineutrinos, our analysis on SN fluxes is not significantly affected by feedback effects. Experiments where electron antineutrinos is not the dominant channel, such as liquid argon experiments (e.g. DUNE \cite{Acciarri:2015uup}) or future very large dark matter detectors (e.g. DARWIN \cite{Aalbers:2016jon}), would be highly sensitive to such feedback effects. Comparison of the signals obtained in different experiments would be a way to learn the utmost from a future galactic supernova.

At the large densities of a supernova core neutrinos are not able to stream freely. The neutrinosphere is defined as the approximate shell from where neutrinos of energy $E_\nu$ can escape without substantial further diffusion. Analytically we can define it as the radial position $R_\nu(E_\nu)$ where the optical depth is unity:
\begin{equation}
\tau_\nu(E_\nu)=\int^\infty_{R_\nu(E_\nu)}\frac{dr}{\lambda(r,E_\nu)}=1.
\label{eq:neutrinosphere}
\end{equation}
Here $\lambda=(\sum_in_{r,i}\sigma_i(E_\nu))^{-1}$ is the mean free path for neutrinos, with $n_{r,i}$ the number density of target particles of species $i$ and $\sigma_i (E_\nu)$  the corresponding interaction cross section with matter. Because of the $E_\nu^2$ dependence of the low-energy scattering cross section on nucleons, there is a separate neutrinosphere for each neutrino energy. 

\begin{figure}[t]
\includegraphics[width=0.49\textwidth]{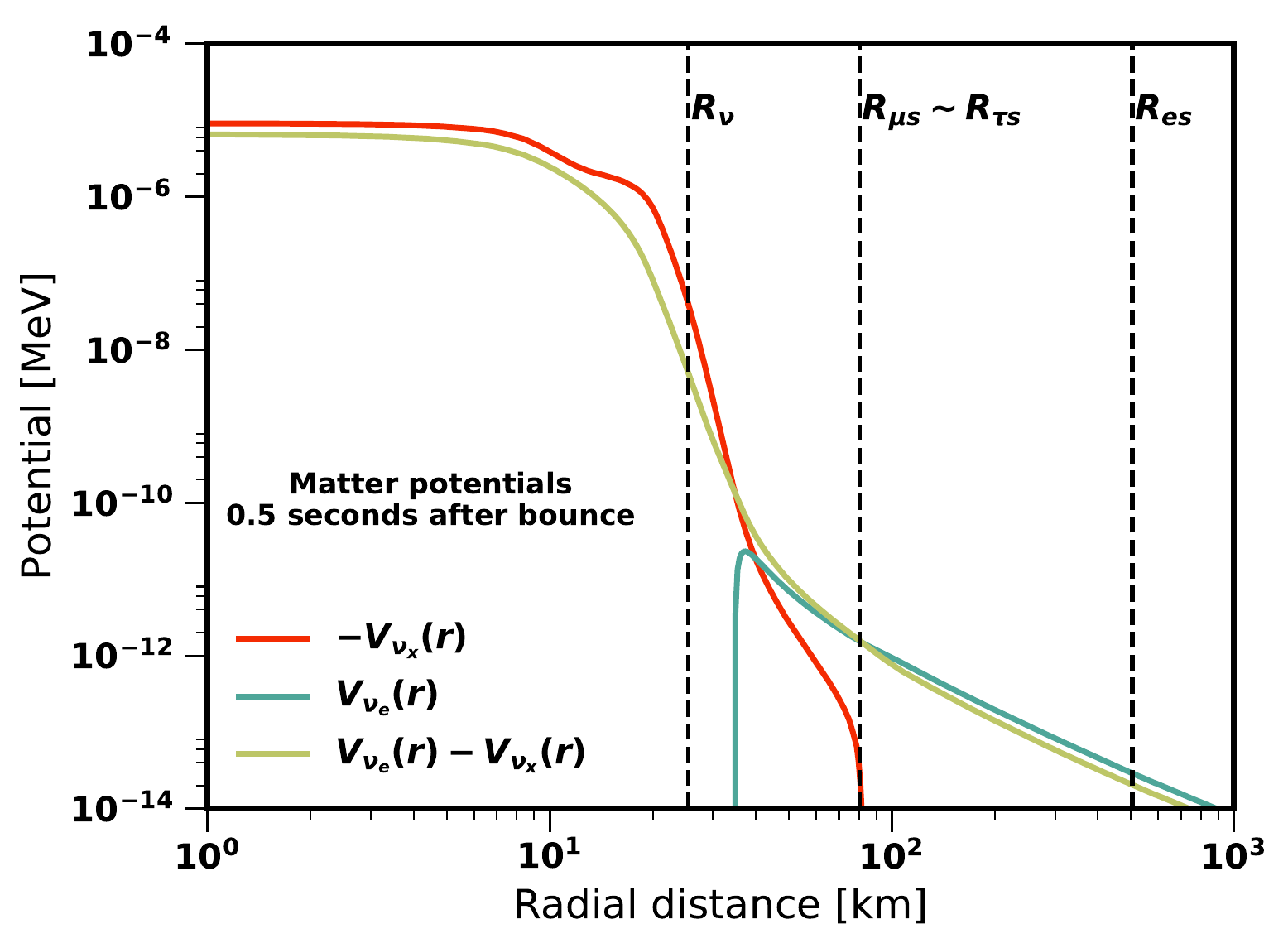}
\caption{Matter potentials half a second after bounce. Vertical lines for the neutrinosphere and the resonance location for the active-sterile conversions assuming $E_\nu=$15 MeV, $\Delta m_{41}^2=$1 eV and $\theta_{14}=\theta_{24}=\theta_{34}=10\degree$. The active-active resonances occur further at lower densities.}
\label{fig:potentials}
\end{figure}

If a resonance occurs inside the neutrinosphere there might be replenishment of active states, making the problem more difficult to treat \cite{Arguelles:2016uwb} and a dedicated investigation is needed. For the parameters under consideration in this work all resonances occur outside the neutrinosphere (see Figure \ref{fig:potentials} for an example). Larger mass splittings correspond to resonances in higher density regions, where the potential is larger. For keV-scale sterile neutrinos, an interesting case for dark matter studies, the resonances associated with active-sterile mixing occur inside the neutrinosphere \cite{Arguelles:2016uwb}.

From equation (\ref{eq:coherence_loss}) we  see that a fraction of the events, related to the heaviest state, will arrive at the detector delayed by:
\begin{equation}
\frac{\Delta L}{c}= \text{5.15 ms}\left(\frac{L}{\text{10 kpc}}\right)\left(\frac{\text{10 MeV}}{E_\nu}\right)^2\left(\frac{\Delta m_{41}}{\text{1 eV}}\right)^2,
\end{equation}
For keV-scale sterile neutrinos this delay is in the order of an hour, in contrast to miliseconds for the eV-scale case. This feature would be useful to determine the mostly sterile state mass, but in our case we neglect it as a millisecond delay is not easily observable.

\section{Detection}
We assume that when the next SN burst is observed in our Galaxy, the largest neutrino detectors will be Hyper-Kamiokande and IceCube. The SN neutrinos travelling through water or ice interact dominantly through inverse beta decay reactions on free protons ($\bar{\nu}_e+p\rightarrow n+e^+$) and Cherenkov light is generated by the secondary positrons, which is then observed by photomultipliers. Free protons means hydrogen nuclei, and not the protons in oxygen, for which nuclear binding effects suppress interactions at low energies.

Event-by-event energy information will be achievable in  Hyper-Kamiokande. IceCube, on the other hand, is designed to reconstruct interactions of neutrinos with energies above $\sim 100$~GeV. The spacing between photomultipliers of 17 m vertically and 125 m horizontally means they cannot reconstruct the Cherenkov rings of individual MeV neutrino interactions. However, having a low dark noise and a huge amount of interactions allows for the detection of $\mathcal{O}$(10 MeV) neutrinos from galactic core collapse supernovae from the collective rise in all photomultiplier rates on top of the dark noise \cite{Halzen:1994xe,Halzen:1995ex,Abbasi:2011ss}.

The total cross section for inverse beta decay is
\begin{equation}
\sigma(E_\nu)=\left[9.52\times 10^{-44}\,\text{cm}^2(E_\nu-1.3\,\text{MeV})^2\right]\left(1-7\frac{E_\nu}{m_p}\right)
\end{equation}
where $m_p$ is the proton mass, and the energy of the detected positron is $E_p = (E_\nu-1.3\,\text{MeV})(1-E_\nu/m_p)$ \cite{Beacom:2010kk}. This approximation is very accurate at low energies ($E_\nu\leq 60$ MeV) thus can be safely used for supernova neutrino analyses \cite{Strumia:2003zx}.

Hyper-Kamiokande is planned to have a total mass around 0.3 to 0.5 Mton \cite{Hyper-Kamiokande:2016dsw}. For a neraby SN explosion we can consider the total mass of the detector as fiducial due to the large number of events in a short amount of time. Icecube's lattice of photomultiplier tubes monitor a volume of approximately a cubic kilometer in the deep Antarctic ice. A fair comparison in terms of statistical accuracy is needed given the different observation method in IceCube. A study of the initial 380 ms of the burst in the Lawrence Livermore 20 $M_\odot$ SN simulation at a distance of 10 kpc would require a 0.45 Mton background free detector to statistically compete with IceCube \cite{Abbasi:2011ss}. 

\begin{figure}[t]
\includegraphics[width=0.49\textwidth]{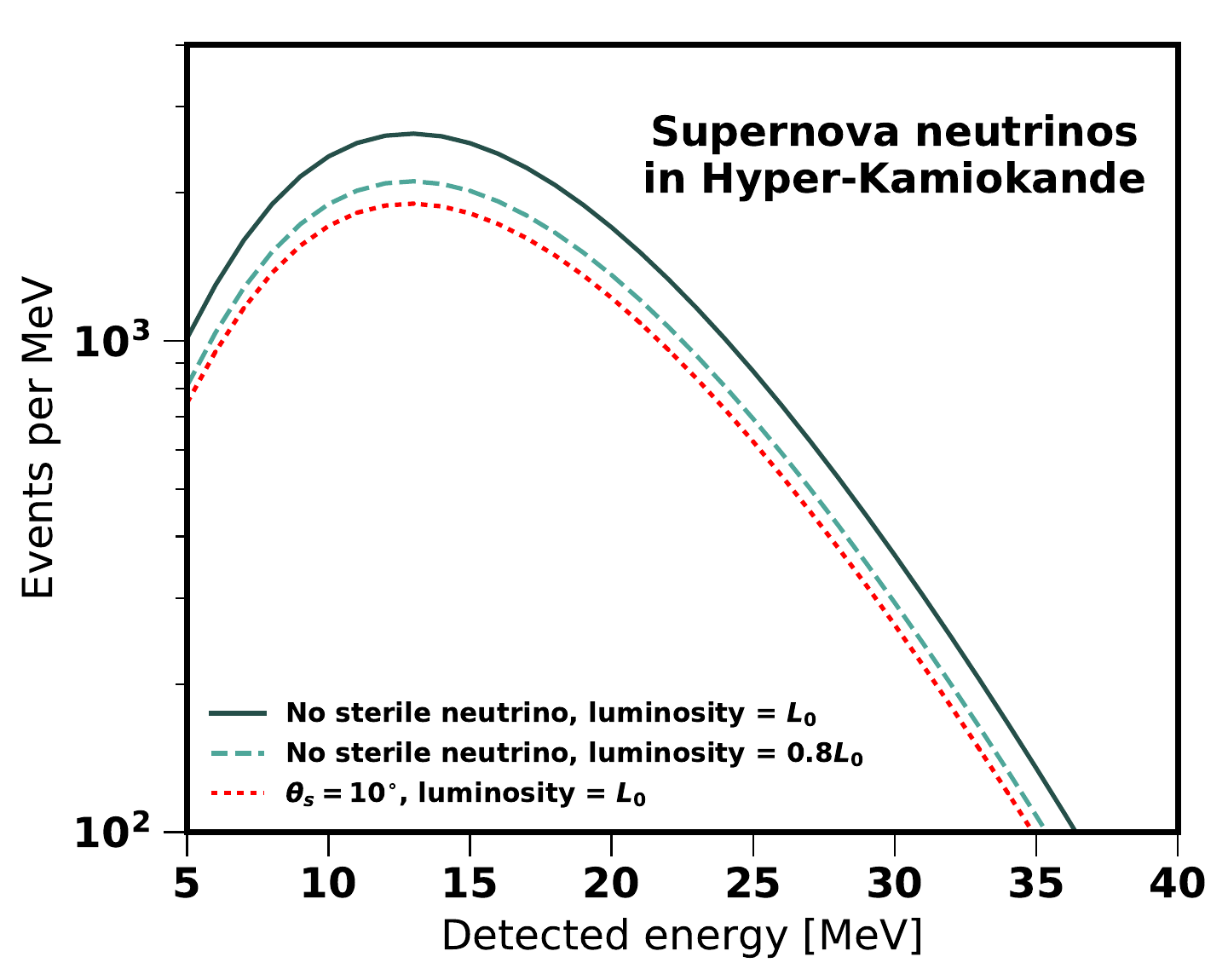}
\caption{Neutrino spectrum in Hyper-Kamiokande from an 8.8 $M_\odot$ SN at a 10 kpc distance. Mixing with a sterile neutrino suppresses the overall flux. This can be compensated by changing the total luminosity of the neutrino burst. For this plot we assume $\Delta m_{41}^2=$1 eV and $\theta_{14}=\theta_{24}=\theta_{34}=10\degree$. }
\label{fig:HK_1}
\end{figure}

Our estimate for the detection rate in IceCube follows ref. \cite{Dighe:2003be}. We have also included DeepCore's 360 photomultipliers with a 35\% larger quantum efficiency \cite{Collaboration:2011ym}. For the aforementioned $8.8 M_\odot$ supernova at 10 kpc Hyper-Kamiokande would observe $\mathcal{O}(10^4)$ events in comparison to $\mathcal{O}(10^5)$ events in IceCube. Nonetheless Hyper-Kamiokande can measure the energy spectrum while IceCube can only tell us the total flux.

An example of the expected neutrino spectrum from a supernova in Hyper-Kamiokande shown in figure \ref{fig:HK_1}. The flux of active neutrinos is reduced due to mixing with the sterile flavour, but this can be compensated by changing the total luminosity of the neutrino burst. Hence we expect that uncertainties in the measurement of the SN flux will introduce systematics into measurements of the sterile neutrino properties.

\section{Results}
\begin{figure*}[t]
\includegraphics[width=0.24\textwidth]{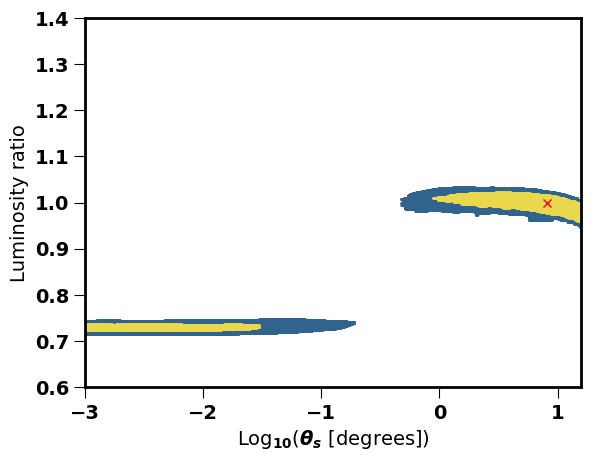}
\includegraphics[width=0.24\textwidth]{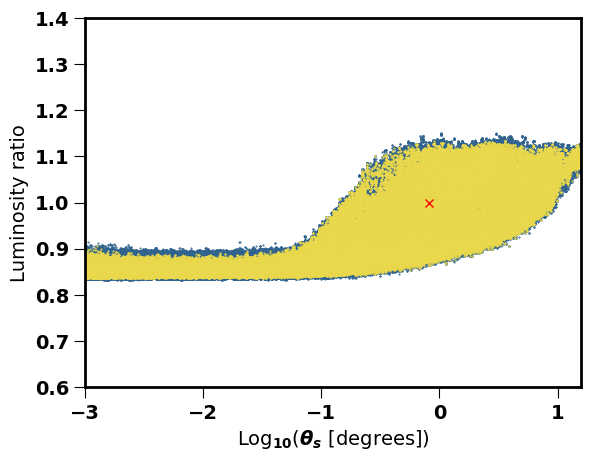}
\includegraphics[width=0.24\textwidth]{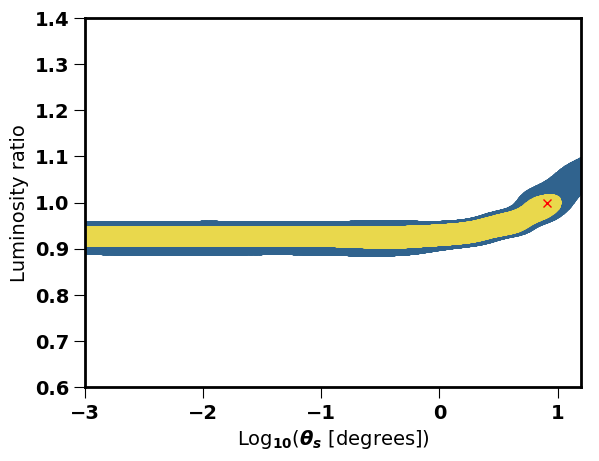}
\includegraphics[width=0.24\textwidth]{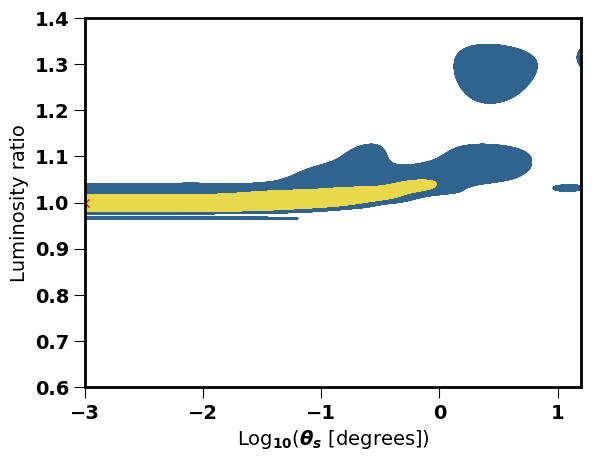}
\caption{Two dimensional posterior distributions at $68\%$ confidence (yellow) and $95\%$ confidence (blue) for Hyper-Kamiokande, comparing the best-fit active-sterile mixing angle to the uncertainty on the supernova neutrino luminosity for scenarios A to D, from left to right. Red points mark the fiducial values for each scenario.  In all cases Hyper-Kamiokande is able to reconstruct the fiducial parameters, but  independent measurements will be required to break the degeneracy between luminosity and mixing angle.}
\label{fig:all_2d}
\end{figure*}

In this section we look at how well Hyper-Kamiokande and IceCube will be able to measure the flux of neutrinos from a galactic supernova. In order to properly account for the degeneracies between the overall flux and the presence of sterile neutrinos in the spectrum, we perform a Markov Chain Monte Carlo (MCMC) analysis over parameters on which the SN flux depends.  In this way we hope to find out if the presence of eV sterile neutrinos can be inferred from the measurements of a 10 kpc distant supernova at Hyper-Kamiokande and IceCube given the uncertainties of the physics of supernovae. For the sterile neutrinos we have two free parameters: the mixing angle $\theta_s$, which is assumed to be the same for every active-sterile mixing ($\theta_s=\theta_{14}=\theta_{24}=\theta_{34}$) and the additional mass splitting $\Delta m_{41}$. For the properties of the supernova itself we consider uncertainties on the total flux and the average neutrino energies. We implement this using two parameters, which are the ratio of each of these quantities to the values obtained in the Garching simulation \cite{Huedepohl:2009wh}. For IceCube we do not scan over the average energy, as the neutrino spectrum can not be measured, but instead we have an additional parameter for the quoted systematic uncertainties for the detector itself.

Each of these parameters has associated with it a prior distribution, which represents how well each of these parameters are known before the analysis is performed. For the sterile neutirno parameters we use log-flat priors, while for the other parameters we use linear flat priors. After the analysis the MCMC combines these priors with the likelihood from the simulated data to give the posterior distribution. This gives us the preference the data has for particular combinations of the four parameters in our fit, and makes it easy to see any degeneracies between these parameters.

Since we do not know whether sterile neutrinos exist, or their properties if they do, we consider different scenarios. Table \ref{tab:scenarios} shows the four scenarios considered in this work, written in terms of the global fit parameters for the controversial neutrino anomalies: $\theta_s=c_1\times 8.13\degree$ and $\Delta m_{41}=c_2\times\sqrt{1.7}$ eV.  It is not important for our analysis if the current hints of sterile neutrino anomalies go away, we merely use them as benchmarks to study the sensitivity that Hyper-Kamiokande and IceCube will have to sterile neutrinos. The four cases we consider, laid out in table \ref{tab:scenarios}, correspond to global fit to short baseline experiments (A), small active-sterile mixing angle (B), small mostly sterile state mass (C) and no mixing with sterile neutrinos (D).

\begin{table}[h]
\centering
\begin{tabular}{c|c|c|c}
Scenario & $c_1$ & $c_2$ & Description  \\ \hline
A & 1 & 1 & global fit to short baseline experiments \\
B & 0.1 & 1 & small active-sterile mixing angle\\
C & 1 & 0.1 & small mostly sterile state mass \\
D & 0 & - & no active-sterile mixing
\end{tabular}
\caption{Sterile neutrino scenarios considered in the MCMC analysis.}
\label{tab:scenarios}
\end{table}

A feature that can be seen in all scenarios is that Hyper-Kamiokande can provide a precise measurement of a nearby supernova temperature. Most of the observed events come from the inverse beta decay, which is sensitive to the average neutrino energy ($\sigma\propto E_{\bar{\nu}_e}^2$).
A comparison of all four scenarios for the sterile neutrino mixing angle and supernova luminosity in shown in figure \ref{fig:all_2d}, with the full results in Appendix \ref{app}.

For scenario A (Figure \ref{fig:corner_HK_NH_A}) the reconstructed supernova flux is bimodal meaning Hyper-Kamiokande will observe a flux of neutrinos which is consistent with two different physics situations. The first is a supernova with an actual lower absolute luminosity, compared with that from simulation. The second situation is that the overall luminosity is suppressed due to adiabatic conversion into sterile neutrinos. A precise determination of the expected flux without active-sterile mixing is needed to break this degeneracy. The mass splitting $\Delta m_{41}$  can be determined to be on the eV scale, and the probability distribution is peaked relatively close to the true value of $\Delta m_{41}=1.3$ eV.  It is difficult to tell between zero mixing angle and $8.13\degree$ mixing angle, but mixing angles between those two values are clearly disfavoured. While all the parameters are not reconstructed brilliantly, the two acceptable regions, with and without mixing, are nicely separated. In particular, the presence of a sterile neutrino would reduce the reconstructed luminosity by nearly 30\% so that if it were possible to estimate the total luminosity in neutrinos at that accuracy, the effects of sterile neutrinos might be relatively easy to see.

Scenario B (figure \ref{fig:corner_HK_NH_B}) has a smaller mixing angle but the same eV scale mass. Again the reconstructed parameters are consistent either with a smaller luminosity and a small mixing angle or a larger luminosity with a larger mixing angle but the best fit cloud is not well separated into two distinct regions.  When one reconstructs the flux in this situation, a very low mixing angle is consistent with a luminosity which is only 15\% different from the true value, hence a better estimate of the supernova luminosity would be required to spot the presence of the sterile neutrino.


Scenario C (figure \ref{fig:corner_HK_NH_C}) corresponds to a lower mass sterile neutrino. It is clear that the fit is only consistent with the correct luminosity if one assumes the right level of mixing, although one would need an estimate of the overall luminosity from another method with roughly 5\% accuracy in order to tell the difference between having a sterile neutrino and no sterile neutrino.  

Scenario D (figure \ref{fig:corner_HK_NH_D}) corresponds to no mixing with sterile neutrinos.  We  place a tight constraint on the presence of sterile neutrinos, again assuming that we have an independent constraint upon the overall luminosity of the supernova explosion.  Hyper-Kamiokande can reconstruct the supernova neutrino luminosity and average energy with only a one or two percent error, which is extremely impressive, but requires knowledge of the sterile neutrino properties. 


What is clear from all figures is that without a precise knowledge of the expected luminosity of the supernova neutrino burst, it is very difficult to place a constraint on the existence of sterile neutrinos. Conversely, without knowing whether sterile neutrinos exist and mix with the active flavours, we can not make an accurate measurement of a supernova neutrino luminosity.

Our analysis of the IceCube data   leads to a less precise measurement of the parameters of the supernova, as shown in figures \ref{fig:corner_IC_NH_A}, \ref{fig:corner_IC_NH_B}, \ref{fig:corner_IC_NH_C} and \ref{fig:corner_IC_NH_D}.  In particular our analysis shows IceCube underestimating the total flux of the supernova by a significant factor so that while it gives an excellent order of magnitude detection of the luminosity, it is not precise enough to exclude or detect the effects of new physics.  We have done an extremely simple analysis and it would be very interesting to see if the IceCube collaboration themselves could do a better job of reconstructing more accurate parameters.

\section{Conclusions}

In this work we have examined how well one can reconstruct various parameters associated with a supernova explosion with and without a sterile neutrino that couples to all the flavours of neutrinos equally.  We estimated how accurately the  Hyper-Kamiokande and IceCube experiments will be able to measure the luminosity and the average temperature of the neutrino flux from a future galactic supernova  burst.

We focused on electron antineutrinos, for which feedback effects can be safely neglected.
This is because their resonance occurs before the electron abundance is altered due to feedback. This is fortuitous, especially since Hyper-Kamiokande is particularly sensitive to electron antineutrinos.

Our main result is that the mixing angle and the mass of the sterile neutrino can be obtained by analysing the spectrum of neutrinos coming from the supernova, provided that the supernova burst is well understood beforehand. This means that a precise knowledge of the expected neutrino luminosity is required to break the degeneracy between active-sterile mixing and supernova luminosity. This is shown in figure \ref{fig:all_2d} for Hyper-Kamiokande. For IceCube the lack of information on the spectrum of the neutrino burst makes the measurement more difficult, resulting in less precise constraints. 

It is extremely interesting to consider which kinds of observations would provide information about the overall neutrino rate, so that these  degeneracies between luminosity and new physics could be broken.  In particular future large dark matter detectors such as DARWIN \cite{Aalbers:2016jon} (and possibly Argo \cite{Aalseth:2017fik}) should be able to detect neutrinos via coherent nuclear scattering.  Since this detection method would be sensitive to the flux of all active neutrinos, it would  provide new information which might constrain sterile neutrino scenarios more effectively. Unfortunately knowledge of the impact of a sterile neutrino within a supernova simulation including feedback effects would be required in order to make reliable estimates of the expected flux of the neutrino species other than antineutrinos. Since we cannot estimate the relevant neutrino fluxes from the existing simulations in a self-consistent way we have been unable to include such detectors in our analysis.

Whatever the complications it is clear from this work that if Hyper-Kamiokande is operating when a supernova goes off in the relatively local Galaxy, we will learn a huge amount about the astrophysics taking place during these incredible events and also at the very least place important constraints on beyond the Standard Model particle physics.

\section*{Acknowledgements}
TF thanks support from CNPq SwB grant. The research leading to these results has been funded by the European Research Council through the project DARKHORIZONS under the European Union's Horizon 2020 program (ERC Grant Agreement no.648680).  The work of MF was also supported partly by the  STFC Grant  ST/L000326/1.

\appendix
\section{MCMC results \label{app}}
\begin{figure*}[t]
\includegraphics[width=0.95\textwidth]{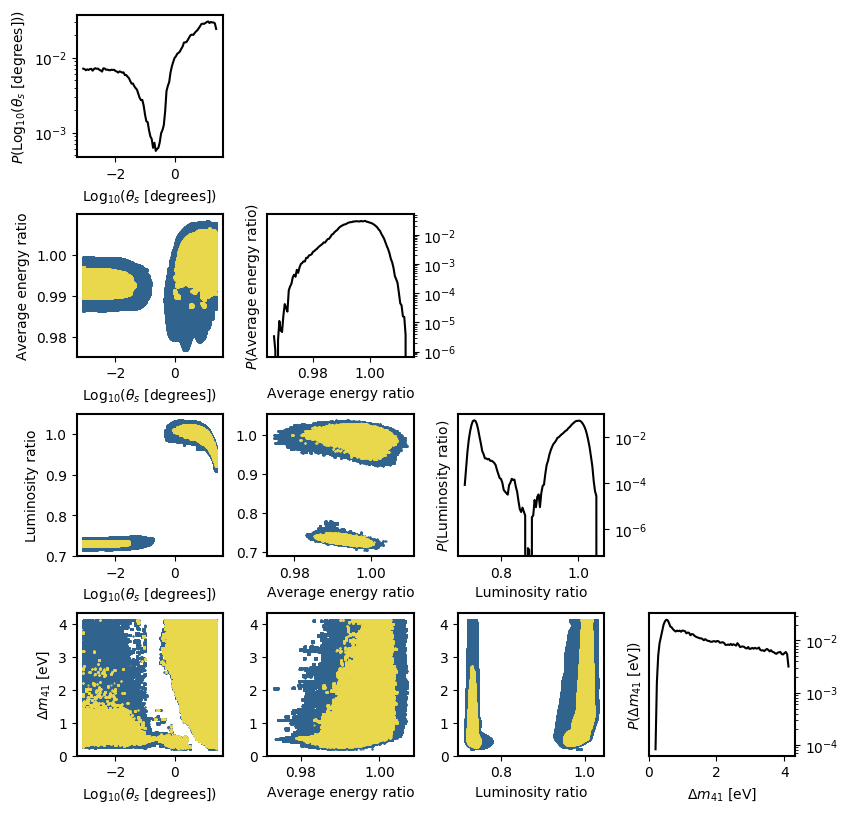}
\caption{Corner plot of the parameters from our MCMC projection for a $d=10$kpc supernova on Hyper-Kamiokande and scenario A ($\theta_s=8.13\degree$, $\Delta m_{41}=1.3\,{\rm eV}$). The contours bound a given fraction of the total integrated posterior (0.68 in yellow and 0.95 in blue). For the one-dimensional histograms, P(x) is the one-dimensional posterior for the parameter x.}
\label{fig:corner_HK_NH_A}
\end{figure*}

\begin{figure*}[t]
\includegraphics[width=0.95\textwidth]{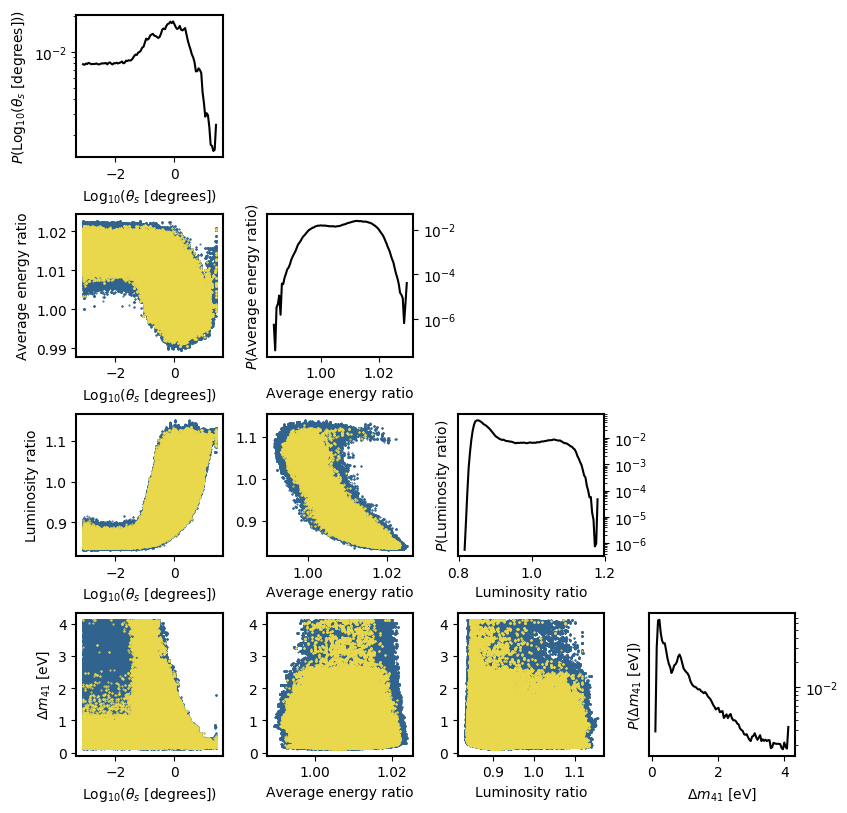}
\caption{Same as Figure \ref{fig:corner_HK_NH_A} for scenario B ($\theta_s=0.813\degree$, $\Delta m_{41}=1.3\,{\rm eV}$).}
\label{fig:corner_HK_NH_B}
\end{figure*}

\begin{figure*}[t]
\includegraphics[width=0.95\textwidth]{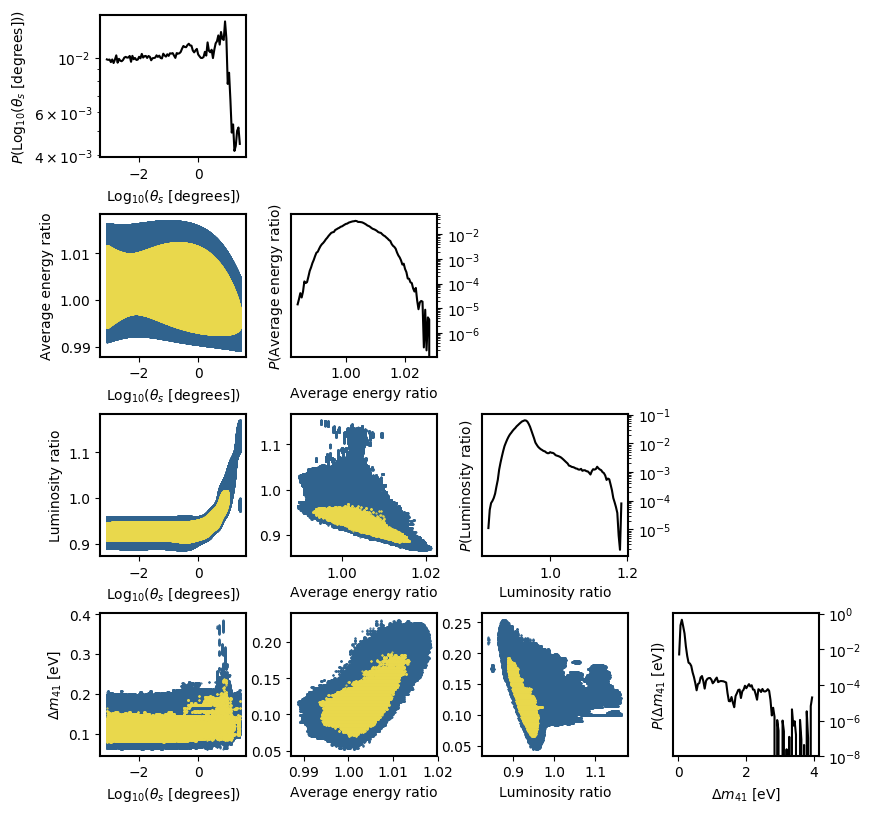}
\caption{Same as Figure \ref{fig:corner_HK_NH_A} for scenario C ($\theta_s=8.13\degree$, $\Delta m_{41}=0.13\,{\rm eV}$).}
\label{fig:corner_HK_NH_C}
\end{figure*}

\begin{figure*}[t]
\includegraphics[width=0.95\textwidth]{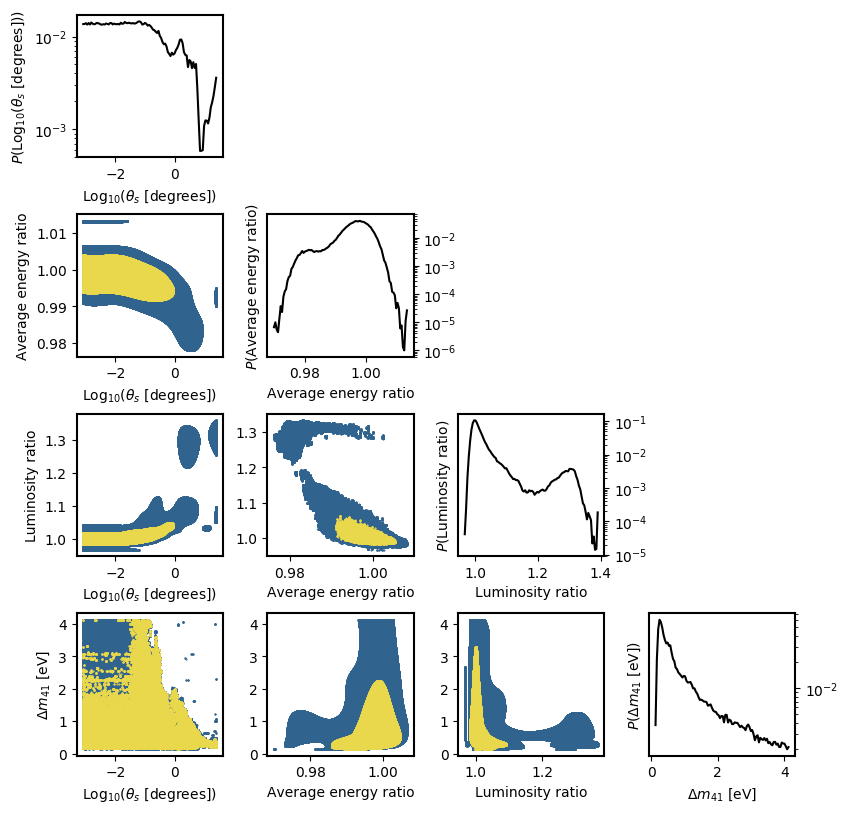}
\caption{Same as Figure \ref{fig:corner_HK_NH_A} for scenario D (no sterile neutrino).}
\label{fig:corner_HK_NH_D}
\end{figure*}

\begin{figure*}[t]
\includegraphics[width=0.95\textwidth]{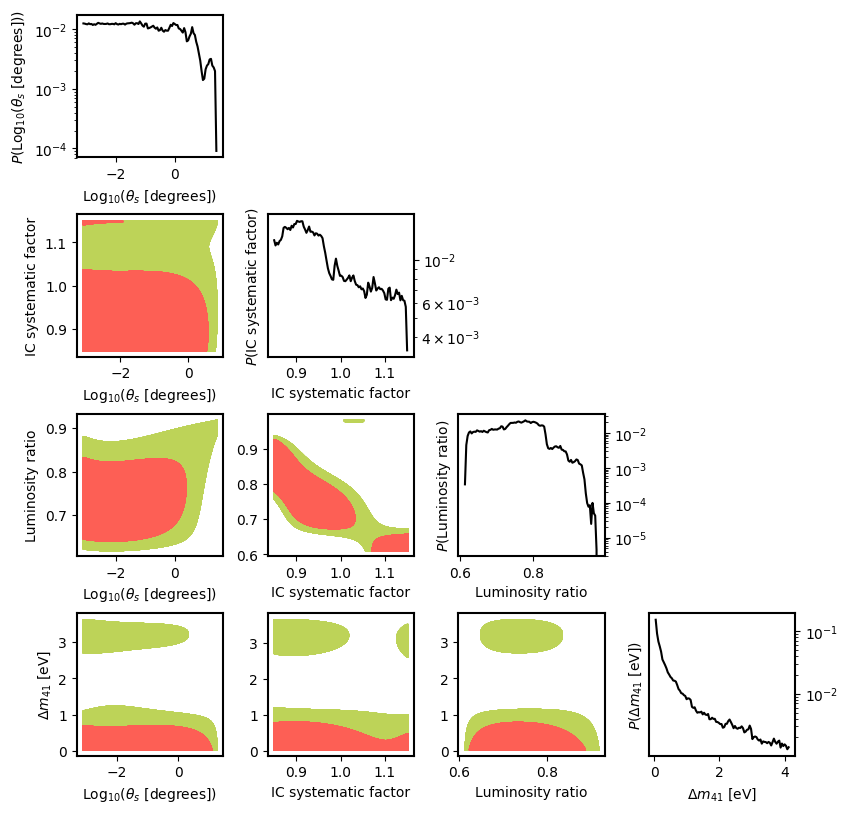}
\caption{Corner plot of the parameters from our MCMC projection for a $d=10$kpc supernova on IceCube and scenario A ($\theta_s=8.13\degree$, $\Delta m_{41}=1.3\,{\rm eV}$). The contours bound a given fraction of the total integrated posterior (0.68 in red and 0.95 in green). For the one-dimensional histograms, P(x) is the one-dimensional posterior for the parameter x.}
\label{fig:corner_IC_NH_A}
\end{figure*}

\begin{figure*}[t]
\includegraphics[width=0.95\textwidth]{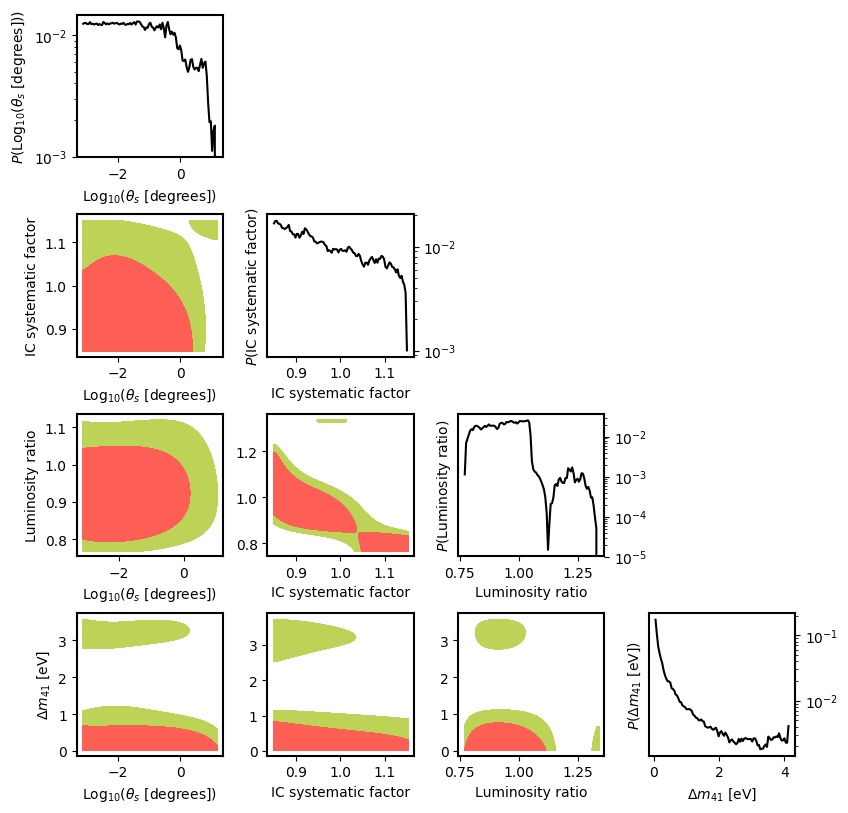}
\caption{Same as Figure \ref{fig:corner_IC_NH_A} for scenario B ($\theta_s=0.813\degree$, $\Delta m_{41}=1.3\,{\rm eV}$).}
\label{fig:corner_IC_NH_B}
\end{figure*}

\begin{figure*}[t]
\includegraphics[width=0.95\textwidth]{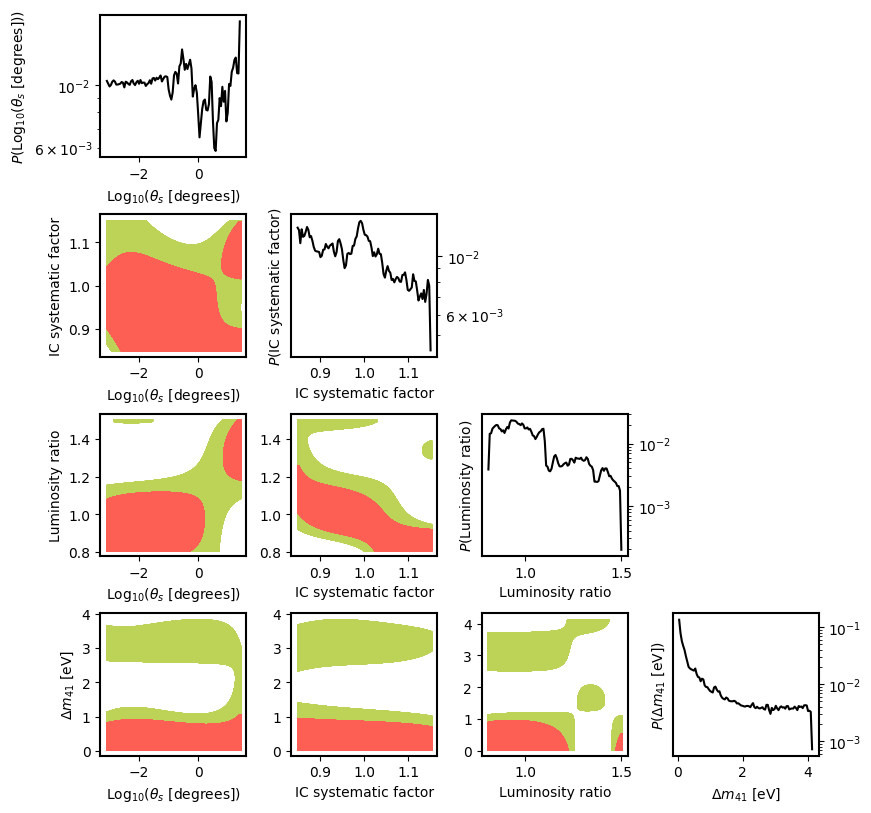}
\caption{Same as Figure \ref{fig:corner_IC_NH_A} for scenario C ($\theta_s=8.13\degree$, $\Delta m_{41}=0.13\,{\rm eV}$).}
\label{fig:corner_IC_NH_C}
\end{figure*}

\begin{figure*}[t]
\includegraphics[width=0.95\textwidth]{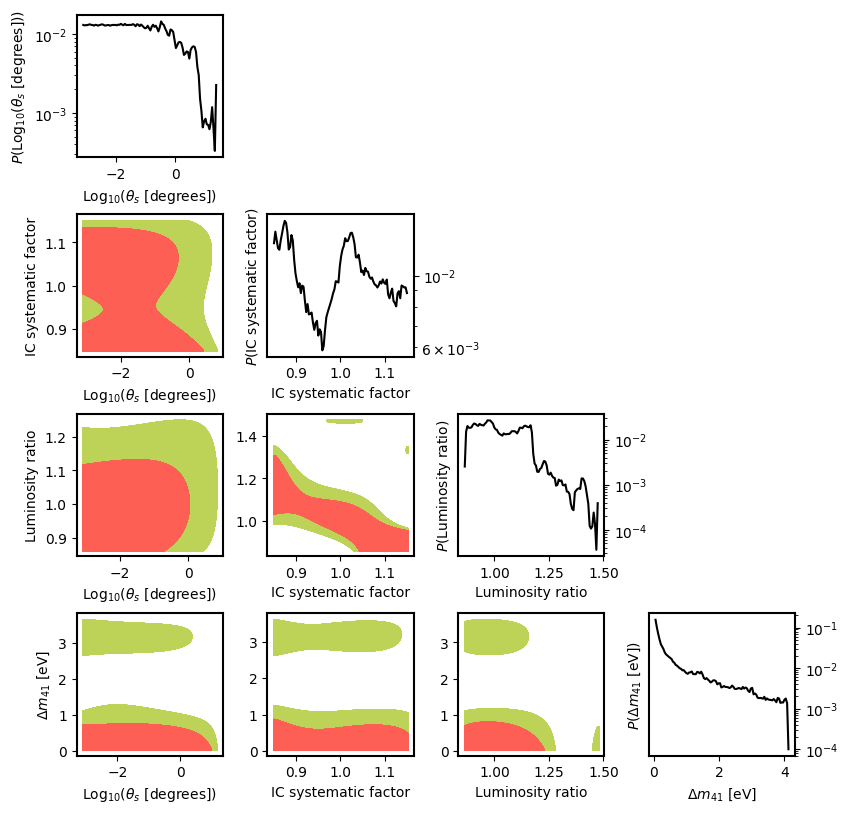}
\caption{Same as Figure \ref{fig:corner_IC_NH_A} for scenario D.}
\label{fig:corner_IC_NH_D}
\end{figure*}

Here we show all of the corner plots for both our Hyper-Kamioknade and IceCube analyses. Each panel shows the posterior marginalised over all other parameters for both the two-dimensional and one-dimensional cases. Hyper-Kamiokande plots are shown in figures \ref{fig:corner_HK_NH_A}, \ref{fig:corner_HK_NH_B}, \ref{fig:corner_HK_NH_C} and \ref{fig:corner_HK_NH_D} for scenarios A to D respectively. IceCube plots are  shown in figures \ref{fig:corner_IC_NH_A}, \ref{fig:corner_IC_NH_B}, \ref{fig:corner_IC_NH_C} and \ref{fig:corner_IC_NH_D} for scenarios A to D respectively.

\end{document}